\documentclass[journal]{IEEEtran}

\usepackage{graphicx}

\ifCLASSINFOpdf
\else
\fi
\begin{document}
%
\title{Large Language Models (LLMs): \\Deployment, Tokenomics and Sustainability}
%
%
%

\author{Haiwei Dong~\IEEEmembership{Senior Member,~IEEE},
        Shuang Xie~\IEEEmembership{Member,~IEEE}
}

%
%

\markboth{CTSoc News on Consumer Technology}%
{Shell \MakeLowercase{\textit{et al.}}: Bare Demo of IEEEtran.cls for IEEE Journals}
%



\maketitle

\begin{abstract}
The rapid advancement of Large Language Models (LLMs) has significantly impacted human-computer interaction, epitomized by the release of GPT-4o, which introduced comprehensive multi-modality capabilities. In this paper, we first explored the deployment strategies, economic considerations, and sustainability challenges associated with the state-of-the-art LLMs. More specifically, we discussed the deployment debate between Retrieval-Augmented Generation (RAG) and fine-tuning, highlighting their respective advantages and limitations. After that, we quantitatively analyzed the requirement of xPUs in training and inference. Additionally, for the tokenomics of LLM services, we examined the balance between performance and cost from the quality of experience (QoE)'s perspective of end users. Lastly, we envisioned the future hybrid architecture of LLM processing and its corresponding sustainability concerns, particularly in the environmental carbon footprint impact. Through these discussions, we provided a comprehensive overview of the operational and strategic considerations essential for the responsible development and deployment of LLMs.

\end{abstract}


%
\IEEEpeerreviewmaketitle

\section{Ubiquitous LLMs}

The recent unveiling of GPT-4o by OpenAI on May 13, 2024 marks a pivotal moment in the evolution of large language models (LLMs) \cite{gpt4-o}. This groundbreaking model, aptly named with ``o” signifying ``omni" for its comprehensive capabilities, transcends the limitations of its predecessors by incorporating multi modality. This signifies a significant step towards achieving more natural and intuitive human-computer interaction.

The emergence of LLMs started from the launch of ChatGPT in November 2022 after two months of which, it reached 100 million monthly users. Since then, many technical companies started to build their open or closed foundation LLM models, such as Gemini (Google), GPT-4 (OpenAI), LLaMA (Meta), Claude (Anthropic), Falcon (TII), Mistral (Mixtral), etc. Initially, text-to-text is the conventional scenario of the LLMs. All questions from users and replied answers from LLMs are words. Starting from 2023, LLMs' capability was extended to multi-modality. In contrast to simple text-to-text mode, the LLMs now support:

\begin{itemize}
    \item Text-to-Image: Generate an image from text descriptions.
    \item Text-to-Video: Generate a video or movie from text descriptions, a bit like story-telling technologies.
    \item Text-to-Audio/Sound: Generate a human-like speech or various environmental sounds from text descriptions.
    \item Image-to-Text: Summarize an image in a few sentences, like generating a caption of an image or photo.
    \item Video-to-Text: Summarize a video in a few sentences, e.g., creating a summary of a YouTube video clip or Netflix movie.
    \item Audio/Sound-to-Text: Describe a speech or surrounding sounds in a few sentences.
\end{itemize}

If we consider 3D objects as a kind of modality, the list of LLMs' capability could even extended in the following:

\begin{itemize}
    \item Text-to-3D: Generate realistic 3D objects from text descriptions.
    \item Image-to-3D: Generate 3D scenes with an image or a 3D avatar with an portrait photo.
    \item Video-to-3D: Generate 3D scenes or eventually an entire 3D virtual environment by videos.
\end{itemize}

It is noted that if it is from low-dimensional modality to high-dimensional modality, it is a decoder/diffusion mode depicted as ``generation". Conversely, it is an encoder mode described as ``summarization".

\begin{figure}[htbp]
\centering
\includegraphics[width=0.47\textwidth]{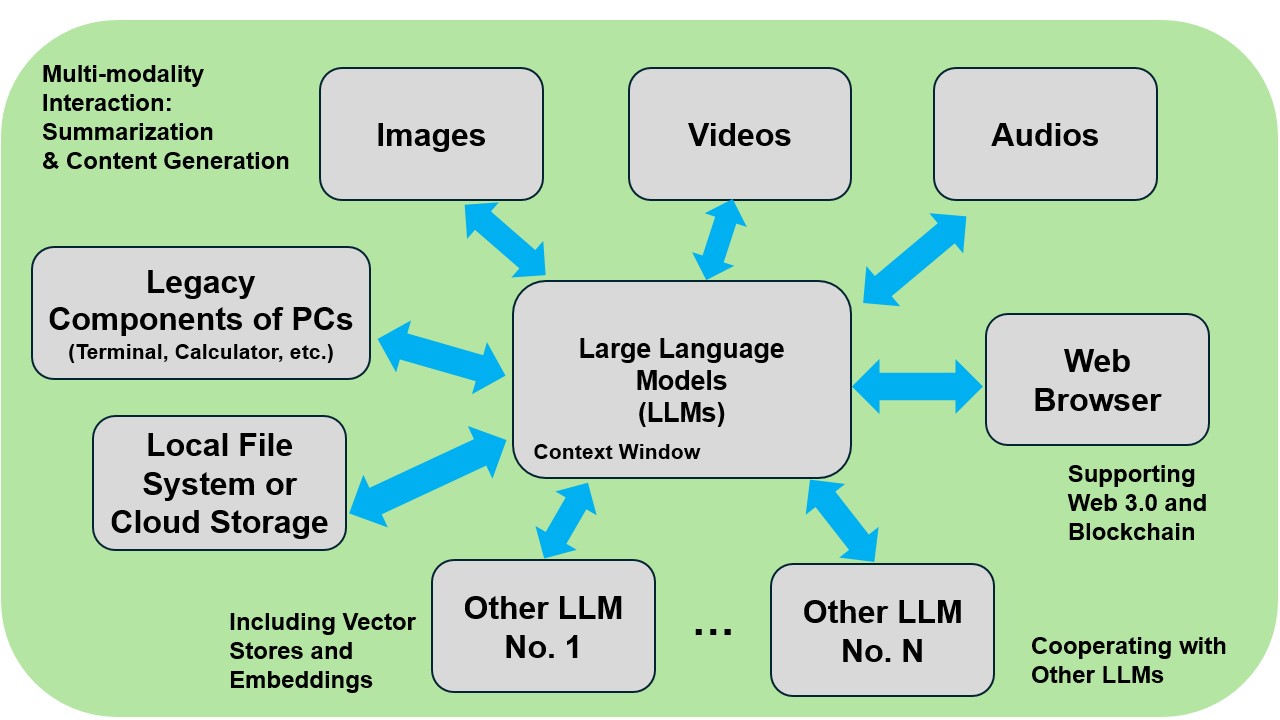}
\caption{LLMs will be the computational core to interact with multimodal multimedia, legacy file system, components of PC, and other LLMs.}
\label{fig:llm_interactions}
\end{figure}

The use cases of LLMs are always as digital assistants in consumer electronics, such as smartphones, laptops, wearable devices, automotive, etc. For example, users speak to their smartwatch about a query which is converted to text by automatic speech recognition (ASR) module, like Whisper. The LLMs running on central/fog/edge cloud answer the query by considering the context of users (e.g., schedule, preference, etc.) with prompt engineering and finally provide a customized natural response by automatic text-to-speech conversion. For eXtended reality (including VR/AR/MR) use case \cite{new_york}, the LLMs not only can generate human-like conversations for avatars, but also generate high-quality point clouds of the 3D objects and their corresponding textures. In general, in our vision, we speculate that LLMs would replace the role of CPU in our traditional computer architecture (shown in Fig. \ref{fig:llm_interactions}) and interact with web browsers, legacy components of PCs, local file systems, various multimedia modalities (including but not limited to images, audios, videos), and other LLM models.

\section{LLM Deployment Debate: RAG vs Fine-tuning}

LLMs are highly effective in a wide range of natural language processing (NLP) tasks \cite{zhao2023survey}, but they often fall short in specific real-world applications due to a lack of specialized knowledge during their training. These models are typically trained on extensive, web-sourced annotated data, which broadens their general applicability but diminishes their accuracy in specialized areas. For instance, BloombergGPT \cite{wu2023bloomberggpt}, which includes a significant proportion of finance-specific data, has shown improved performance in financial contexts, illustrating the benefits of targeted data training. However, assembling large, clean datasets for specialized training involves considerable logistical and cost challenges. In this case, deploying LLMs that have been trained primarily with generic datasets can lead to challenges in processing long, complex texts, thereby limiting their effectiveness in detailed domain discussions and increasing the likelihood of generating hallucinations and producing irrelevant or incorrect content \cite{vaswani2017attention}. To address this challenge, two main deployment strategies have emerged: Retrieval-Augmented Generation (RAG) and fine-tuning.

\subsection{RAG: Solving Context Window Limitation}

The RAG model enhances LLMs by integrating the external knowledge base to efficiently retrieve relevant information that complements the LLM's inherent knowledge. This retrieved information is then incorporated into the LLM's prompt, providing essential contexts. The RAG mechanism is particularly beneficial for tasks that require factual retrieval or similarity searches, as it directly supplies relevant context, thereby reducing the likelihood of generating inaccurate or hallucinations. RAG can also be further enhanced by integrating techniques for in-context examples' retrieval during runtime. This method involves fetching relevant few-shot examples that are semantically similar to the user's query.

\begin{figure}[htbp!]
\centering
\includegraphics[width=0.5\textwidth]{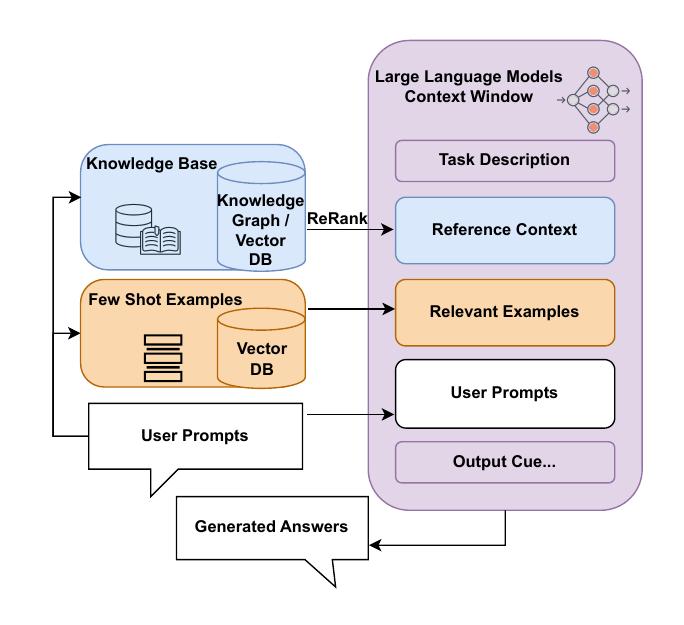}
\caption{Retrieval-Augmented Generation (RAG) architecture for large language models. RAG retrieves relevant data from a knowledge base and optionally few-shot examples to address context window limitations and improve response generation.}
\label{fig:rag}
\end{figure}


An illustration of a RAG-based LLM deployment system is depicted in Fig. \ref{fig:rag}. This system utilizes a knowledge graph database to extract external knowledge and a vector database for retrieving relevant examples. To obtain reference context from the knowledge graph database, the LLM first generates Cypher queries based on the user's prompts and the fields within the Resource Description Framework (RDF) knowledge graph. These queries are then executed in specific knowledge graph databases, such as Neo4j or SPARQL. The retrieved context may need to be re-ranked to ensure it aligns with specific business requirements. For retrieving few-shot examples, which are represented as vectors in a high-dimensional space, the user's query is transformed into the corresponding embedding space. A similarity search is then performed to select the top-n semantically close examples. The final prompt integrates the aforementioned information with the initial user prompt, enabling the LLM to generate responses that are both informative and accurate.

\subsection{Fine-tuning: Deeper Domain Expertise}
Fine-tuning focuses on adapting a pre-trained LLM to a specific task through further supervised learning. This involves adjusting the weights of a pre-trained model on a smaller dataset of task-specific examples. Compared to training a model from scratch, fine-tuning offers significant advantages in terms of computational efficiency and leveraging the pre-trained model's general language understanding.

\begin{figure}[htbp!]
\centering
\includegraphics[width=0.51\textwidth]{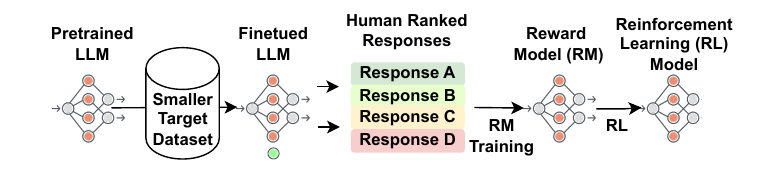}
\caption{LLM fine-tuning for a specific task. The pre-trained model has learned parameters from a massive dataset, while the parameters are updated through fine-tuning in a smaller target dataset. The red dots are the learned parameters in the original models, while the green dots are the new parameters learned from LoRA training. RLHF aligns the model for optimal performance.}
\label{fig:finetune}
\end{figure}

Several fine-tuning techniques have been developed to enhance LLM performance on specific tasks. These include instruction tuning which provides the LLM with explicit instructions or constraints alongside the training data. Low-Rank Adaptation (LoRA) which introduces a small adapter module on top of the pre-trained LLM (shown in Fig. \ref{fig:finetune}). This adapter module is specifically trained on the task-specific data, enabling efficient adaptation without modifying the core LLM architecture \cite{hu2021lora}. Recent advancements in fine-tuning incorporate techniques from Reinforcement Learning from Human Feedback (RLHF) to achieve better alignment between the model's outputs and human expectations \cite{ouyang2022training}. RLHF allows the model to learn through interactions with a human in the loop, who provides feedback on the model's outputs. This feedback is then used to refine the model's parameters and improve its performance on the target task.

\subsection{Takeaway Message}
This section highlights the key considerations when choosing between RAG and fine-tuning for LLM applications. RAG offers faster adaptation and leverages external knowledge bases for factual retrieval and similarity search. However, the quality of the retrieved information heavily relies on the underlying knowledge base. If the knowledge base is outdated, inaccurate, or incomplete, the retrieved information can mislead the LLM to generate factually incorrect outputs. Additionally, RAG struggles with tasks requiring complex reasoning or inference beyond the information explicitly contained in the knowledge base \cite{chen2024benchmarking}. Integrating and reasoning over information from multiple sources can also be challenging for RAG models, potentially leading to inconsistencies or inaccuracies in the generated response. One approach to alleviate these limitations is to employ reranking techniques. Reranking involves applying a secondary sorting step to the retrieved information from the knowledge base. This step can be based on various factors, such as using a confidence score assigned by the retrieval model \cite{cao2018retrieve}, or evaluating whether the retrieved items collectively address all aspects of the user query.

Fine-tuning, on the other hand, provides superior performance for tasks requiring deep domain expertise \cite{balaguer2024rag}. However, it typically requires large, high-quality datasets to achieve optimal performance whereas creating such datasets can be expensive and time-consuming. To address this challenge, some researchers are exploring the use of teacher-student learning. In this approach, a large, pre-trained model (teacher) like GPT4 \cite{gpt4} is used to generate synthetic data that can be used to fine-tune a smaller, open-source model (student) like LLaMA 8B \cite{peng2023instruction}. This approach can be significantly cheaper than creating high-quality human-annotated datasets, especially for tasks where large amounts of data are required. However, it is important to ensure that the synthetic data accurately reflects the real-world distribution of data and that the teacher model itself is not biased. Another challenge is the computational cost of fine-tuning. Fine-tuning often requires significant computational resources, especially for large LLMs. This can be a barrier for smaller organizations or researchers with limited access to computational resources.

The choice between RAG and fine-tuning depends on the specific needs of LLM applications. RAG offers a faster and more adaptable solution for retrieval-based tasks, while fine-tuning provides superior performance for tasks demanding deeper domain expertise.

\section{Training and Inference by xPUs: CPU is Out-of-the-Date?}
The training of LLMs always consumes a lot of computational resources, particularly during the training of foundation models. The referred computational resources herein are xPUs typically including GPUs (graphical processing units), TPUs (tensor processing units), NPUs (neural processing units), LPUs (language processing units), and CPUs (central processing units). The former four are designed for tensor processing (i.e., vector manipulation and calculation) which are composed of tens of thousands of cores. For instance, the newly-release heavy-duty Blackwell GPU from Nvidia, GB202 GPU, has  24,576 CUDA cores. Even the light-duty Nvidia GeForce RTX 4090 has 16,384 CUDA cores. In contrast, CPUs only have very few cores which are designed for sequential tasks rather than parallel matrix or vector calculations. For example, Intel i9-14900 14th Gen has 24 cores. 

There is no doubt that GPUs/TPUs/NPUs/LPUs are the only options to train LLMs. In this case, thousands of xPU cards are occupied to train an LLM for several months. Take the LLaMA 65B model as an example, it has 65 billion parameters. If each parameter is represented by a floating point number, i.e., 2 bytes to be stored, the total required storage would be 130 billion bytes, i.e., 130 gigabytes. As it is more than the memory limit of any single xPU card, the model parameters need to be distributed in different xPUs, named model parallelism. To further improve the training efficiency and training completion time, the training data is shared across xPU cards called data parallelism. Specifically, if we use Nvidia A100 GPU (80GB memory) cards to train the model, considering the overall dataset contains 1.4 trillion tokens (1 token is approximately 0.75 English words), it takes 2048 A100 GPUs for approximately 21 days to complete the training. Even in the fine-tuning phase where domain knowledge tokens are from a few thousand to tens of thousands, 8 GPU cards are always at least required.

 Certainly, the LLM inference can be done by GPUs/TPUs/NPUs/LPUs. For some service providers, the strategy is to use a big number of small xPU cards (in the perspective of memory) to conduct inference. For example, still for LLaMA 65B model, if each parameter is stored in the format of int8 that is one byte, the total required storage is 65 gigabytes. If we use Groq LPU cards in this inference task, each Groq card has 230 megabytes memory. Therefore, the approximate number of Groq cards for LLaMA 65B model inference is (65$\times$1000$\times$1000$\times$1000)$\div$(230$\times$1000$\times$1000) = 282.6 $\doteq$ 283.
 
 However,  these high-end xPU cards are not the only ones that fit the LLM inference scenarios. The main reason is that the inference models are always optimized to be much smaller models in the perspective of parameter number. In addition, each parameter can be used even less accurate format to represent it. In that case, CPUs are also capable of conducting inference without much loss of accuracy. Furthermore, CPUs are the ideal hardware to fulfill the tasks of essential LLM pre-processing, such as embedding and tokenization. Finally, the infrastructures of most existing data centers are composed of CPU servers. By considering the expensive cost of upgrading to high-end AI data centers, the service providers would prefer cost-effective yet usable CPUs for the inference business.

\begin{figure*}[htbp!]
\centering
\includegraphics[width=0.9\textwidth]{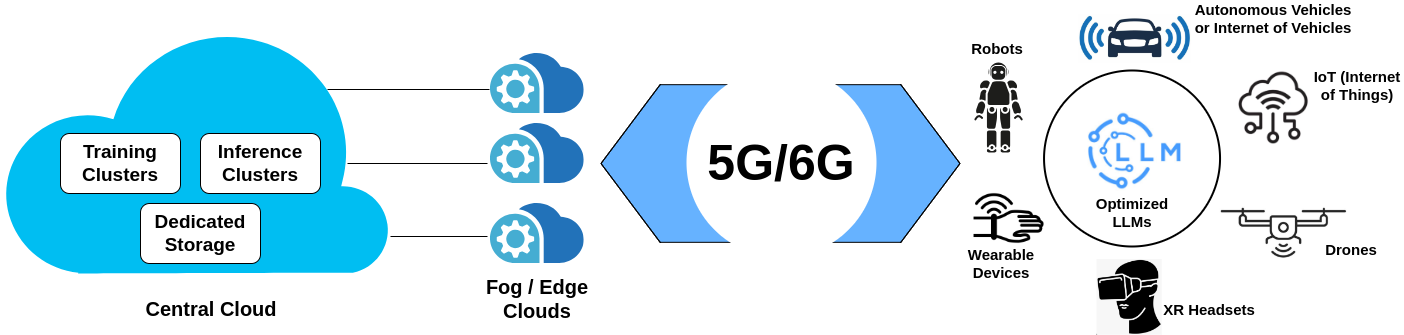}
\caption{The hybrid architecture of AI processing of LLMs. The central and edge clouds and devices work together to deliver high QoE LLM service by balancing factors, including inference accuracy, latency, device capacity, privacy, and security.}
\label{fig:hybrid_llms}
\end{figure*}

\section{Tokenomics vs. Quality of Experience: The Compromise of Performance and Cost}
Tokenomics is a compound word for token and economics referring to the analysis of generative tokens in LLM inference from the perspective of economics. Here, we usually consider two aspects: throughput (tokens per second) and price (USD per 1 million tokens). According to the up-to-date data \cite{tokenomics}, the throughput of most unicorn companies (such as Mistral, Perplexity, Toghether.ai, Anyscale, Deepinfra, Fireworks, Groq, Leption) lies about 50 to 200 tokens per second whereas Groq leads the benchmark to be more than 400 tokens per second. In regard to prices, most of the aforementioned companies can achieve between \$0.2 USD to \$1.0 USD per 1 million tokens. Perplexity and Groq lead the board herein around \$0.25 USD.

If we consider the LLM inference as a service, the customers are users who ask questions by prompts and receive answers in a paid (e.g., OpenAI GPT-4) or unpaid way (e.g., OpenAI GPT-3.5 or GPT-4o). Each data center with thousands of or even tens of thousands of AI servers is like a factory whose products are tokens. During this service, the metrics to evaluate the quality of experience (QoE) of end users include \cite{qoe_metrics}:
\begin{itemize}
    \item Time to first token (TTFT): The waiting time of the users to receive the response after entering their query. Typically, one second of TTFT is considered to be good enough.
    \item Time per output token (TPOT): The time to receive each generated token by the users. If we consider a TPOT as 100 milliseconds per token, equivalent to 10 tokens per second, leading to 600 words per minute that is faster than most of the users can read.
    \item Latency: The overall time of receiving an answer composed of tokens after the initial query from the users. To be more precise, it can be described as $Latency=TTFT+TOPT*(number\;of\;total\;generated\;tokens)$.
    \item Throughput: The generated tokens per second. It can be considered as a median value in a time window considering applicable queries and users.
\end{itemize}

It is important for users to receive high QoE while using the LLM service. However, better service requires more computational and networking resources. It is speculated that the LLM service will be provided with different levels relating to various subscription options or an one-time charge from users. It would be finally a compromise decision of users in the consideration of performance and cost.

\section{Hybrid LLM: Gravitating towards the Edge}
In our vision, the training of the LLMs happens in central clouds or more specifically AI data centers that are composed of training clusters, inference clusters, and dedicated storage, as shown in Fig. \ref{fig:hybrid_llms}. The models in the AI data centers are the full model without any approximation, such as weight pruning and neural pruning. In order to reduce end-to-end latency and operation costs, the AI processing of LLM inference would be as close to the users as possible in a hierarchy, from central clouds to edge clouds, until the devices. The potential devices to provide LLM service include autonomous vehicles / IoVs (Internet of Vehicles), IoT (Internet of Things), Drones, XR (eXtended Reality) headsets \cite{metaverse_ce,metaverse_resource_allocation}, Wearable devices (e.g., smartwatch), robots (e.g., humanoid robot assistant) \cite{LLM_robot}, etc.  Due to the computational capability of the devices, the LLM models running on the device would be optimized: a much smaller model size (7 to 10 times) while keeping the accuracy satisfied. In practice, the LLMs on the device generate a few tokens (typically four for example) as ``drafts" which are then sent to the cloud for checking the accuracy. Once confirmed or corrected by the cloud LLMs, the speculative decoding process is complete and users receive the LLM answers.

In regard to communication, the xPU servers are connected by ToR (Top-of-Rack) Switches (electrical, optical, or hybrid) which are further connected together according to a specific topology, such as fat-tree, dragonfly, etc. via InfiniBand or Ethernet. Communication between central clouds and fog/edge clouds is WAN (Wide Area Network) which might also involve DCI (Data Center Interconnect) if a training task is distributed in multiple clouds. Between edge clouds and user devices, we believe wireless 5G or 6G in the future would be the most convenient and efficient way. Due to cybersecurity factors, like privacy and security requirements \cite{hybrid_ai}, all packets need to go through firewalls and be encrypted. 

\section{Carbon Footprint and Sustainability of LLMs}
The carbon footprint of a LLM comprises two fundamental components: the operational footprint and the embodied footprint \cite{gupta2021chasing}. The operational footprint encompasses emissions stemming from the energy consumption of the hardware used during pre-training, fine-tuning, and inference. The embodied footprint encapsulates the lifecycle emissions associated with hardware manufacturing, including material extraction, processing, and transportation. Accurately estimating the carbon footprint of LLMs before training is crucial for promoting environmentally conscious development practices. This enables researchers and developers to make informed decisions regarding model design and training procedures, promoting the development of more sustainable LLMs. Tools like mlco2 \cite{lacoste2019quantifying} offer a preliminary assessment based on GPU usage, but they often have limitations, such as being unable to account for dense or mixture-of-experts (MoE) architectures. LLMCarbon \cite{faiz2023llmcarbon}, a recently introduced end-to-end carbon footprint projection model, addresses these shortcomings by providing more comprehensive and nuanced estimations for various LLM architectures, including dense and MoE models. For instance, LLMCarbon estimates that training a GPT-3 model could generate around 553.87 $tCO_{2}eq$ (tonnes of $CO_{2}$ equivalent), compared to actual data, the disparity is only +0.32\% with the actual emit is 536.69 $tCO_{2}eq$. However, the training operational carbon footprint estimation made by mlco2 is 69\% higher than the actual, because mlco2 assumes all devices consistently operate at the peak computing throughput using the peak power. 

The sustainability of LLMs can be viewed through a two-fold lens: economic and environmental. LLMs achieve economic sustainability if the value they generate for organizations (e.g., enhanced efficiency, and improved customer service) exceeds the costs incurred from training, inference, and hardware maintenance. Environmental sustainability requires a multifaceted approach, encompassing renewable energy sources for data centers, energy-efficient model architectures, and hardware designed for low-power AI workloads. By prioritizing both economic and environmental considerations, LLM development can become a powerful force for positive change, driving innovation while minimizing its environmental impact.

\begin{IEEEbiography}[{\includegraphics[width=1in,height=1.25in,clip,keepaspectratio]{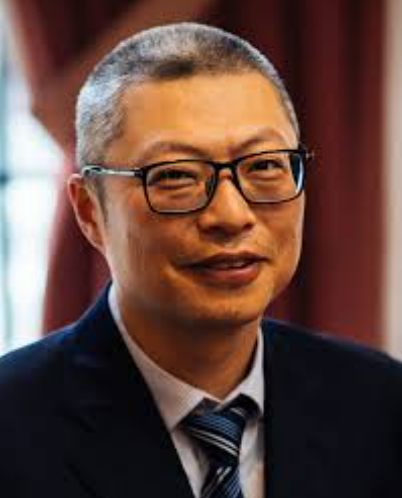}}]{Haiwei Dong}
(haiwei.dong@ieee.org) is currently a Director and Principal Researcher with Huawei Canada, and an Adjunct Professor with the University of Ottawa. He was a Principal Engineer with Artificial Intelligence Competency Center, Huawei Technologies Canada, Toronto, ON, Canada, a Research Scientist with the University of Ottawa, Ottawa, ON, Canada, a Postdoctoral Fellow with New York University, New York City, NY, USA, a Research Associate with the University of Toronto, Toronto, ON, Canada, and a Research Fellow (PD) with the Japan Society for the Promotion of Science, Tokyo, Japan. He received the Ph.D. degree from Kobe University, Kobe, Japan in 2010 and the M.Eng. degree from Shanghai Jiao Tong University, Shanghai, China, in 2008. His research interests include artificial intelligence, multimedia, metaverse, and robotics. He also serves as a Column Editor of IEEE Multimedia Magazine; an Associate Editor of ACM Transactions on Multimedia Computing, Communications, and Applications; and an Associate Editor of IEEE Consumer Electronics Magazine. He is a Senior Member of IEEE, a Senior Member of ACM, and a registered Professional Engineer in Ontario.
\end{IEEEbiography}

\vskip -2\baselineskip

\begin{IEEEbiography}
[{\includegraphics[width=1in,height=1.25in,clip,keepaspectratio]{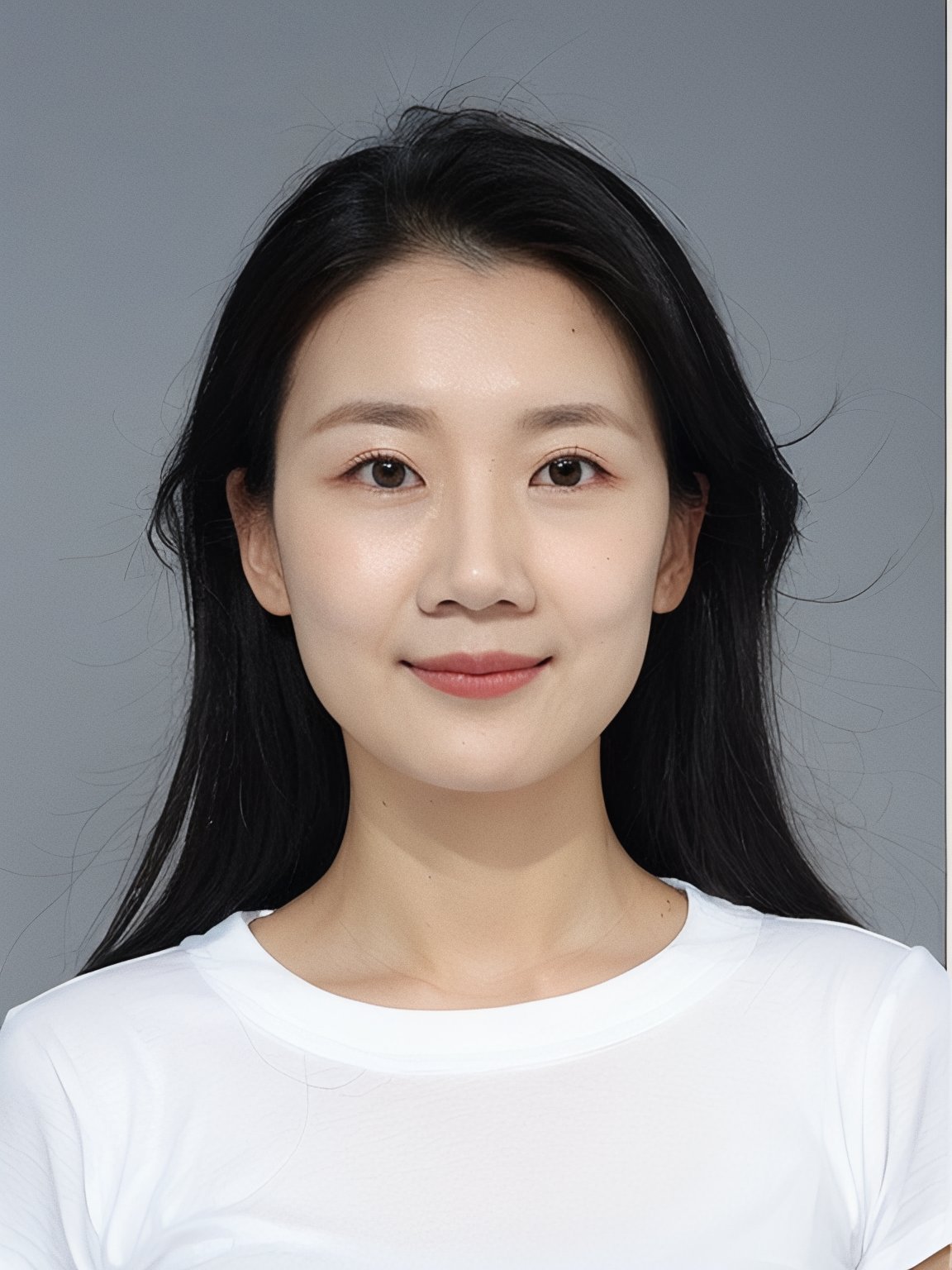}}]{Shuang Xie}
(shuang.xie@ieee.org) is currently a senior Machine Learning Engineer at Shopify, Canada. Her research interests include artificial intelligence, large language models, and computer vision. She received the Master degree from the University of Ottawa, Ottawa, Canada, 2019. Prior to that, she received the M.Eng. degree from Sichuan University, Sichuan, China, 2017. She is a Member of IEEE, and a Member of IEEE Women in Engineering.
\end{IEEEbiography}



\bibliographystyle{IEEEtran}
\bibliography{ref.bib}

\end{document}